\documentclass[aip,amsmath,amssymb,onecolumn]{revtex4-1}
\usepackage[T1]{fontenc}
\usepackage[utf8]{inputenc}
\usepackage{graphicx, color}
\usepackage{amssymb, amsmath, amsfonts, amsthm}
\usepackage{bm}
\usepackage{ifthen}
\usepackage{mathrsfs}
\usepackage{hyperref}
\usepackage{csquotes}

\newcommand{\f}[2]{\mathchoice%
			{\dfrac{#1}{#2}}
	    	{\dfrac{#1}{#2}}
			{\frac{#1}{#2}}
			{\frac{#1}{#2}}}

\newcommand{\ddf}[3][]{\ifthenelse{\equal{#1}{}}{\ensuremath{\f{\dd#2}{\dd#3}}}
{\ensuremath{\f{\dd^{#1}#2}{\dd{#3}^{#1}}}}}
\newcommand{\Dp}[3][]{\ifthenelse{\equal{#1}{}}{\ensuremath{\f{\partial#2}{\partial#3}}}
  {\ensuremath{\f{\partial^{#1}#2}{\partial{#3}^{#1}}}}}
\newcommand{\vect}[1]{\bm{#1}}

\renewcommand{\hat}[1]{\ensuremath{\widehat{#1}}}

\renewcommand{\geq}{\ensuremath{\geqslant}}

\newcommand{\avg}[1]{\ensuremath{\left\langle #1 \right\rangle}}
\newcommand{\expec}[1]{\ensuremath{{\rm I\kern-.3em \rm E}\left[#1\right]}}

\newcommand{\sprod}[2]{\langle#1,#2\rangle}

\newcommand{\e}[1]{\ensuremath{{}_{\text{#1}}}}

\renewcommand{\rm}[1]{\mathrm{#1}}%Pour la compatibilite avec amsmath
\newcommand{\T}{\rm{T}}

\newcommand{\ie}{\textit{i}.\textit{e}. }
\newcommand{\eg}{\textit{e}.\textit{g}. }

\newcommand{\R}{\mathbb{R}} % Les nombres reels
\newcommand{\dd}{\mathrm{d}} % L'\'el\'ement infinit\'esimal.

 \newcommand{\po}{\left(}
\newcommand{\pf}{\right)}

\newcommand{\highlightrevision}{false}

\ifthenelse{\equal{\highlightrevision}{true}}
{
\newcommand{\revision}[1]{{\color{red}{#1}}}
\newcommand{\revbar}[1]{{\color{red}{\sout{#1}}}}
}
{
\newcommand{\revision}[1]{{\color{black}{#1}}}
\newcommand{\revbar}[1]{}
}

\begin{document}

\title[Position-dependent memory kernel in generalized Langevin equations]{Position-dependent memory kernel in generalized Langevin equations: theory and numerical estimation}
\author{Hadrien Vroylandt}
\affiliation{Sorbonne Université, Institut des sciences du calcul et des données, ISCD, F-75005 Paris, France}
\author{Pierre Monmarché}
\affiliation{Sorbonne Université,  Laboratoire Jacques-Louis Lions, LJLL, F-75005 Paris}
\affiliation{Sorbonne Université,  Laboratoire de Chimie Théorique, LCT, F-75005 Paris}

\begin{abstract}
Generalized Langevin equations with non-linear forces and position-dependent linear friction memory kernels, such as commonly used to describe the effective dynamics of coarse-grained variables in molecular dynamics, are rigorously derived within the Mori-Zwanzig formalism. A fluctuation-dissipation theorem relating the properties of the noise to the memory kernel is shown. The derivation also yields Volterra-type equations for the kernel, which can be used for a numerical parametrization of the model from all-atom simulations. 
\end{abstract}

\maketitle

\section{Introduction}

In molecular dynamics simulations, coarse-graining methods consist in replacing an all-atom system by a system with a smaller dimension, either by merging groups of atoms in single units or by describing the system through a relatively small set of collective variables. Such models are then used either to get some physical understanding of the system or, from a numerical viewpoint, to reduce the cost of sampling long trajectories.  In both cases, an important issue is to describe the dynamics of the coarse-grained variables: in theory, obtaining their exact trajectory requires the simulation of the initial all-atom system, which one wants to avoid. In practice, it is approximated by  so-called effective dynamics, namely closed equations which do not involve explicitly the withdrawn atomistic degrees of freedom (whose influence may be taken into account as a random noise) \cite{Peter2009,Lei2010,Legoll2010,Zhang2017,Schilling2021}.

A class of models frequently used to perform this task is given by the Generalized Langevin Equations (GLE) \cite{Zwanzig2001,Chorin2000,Chorin2002,Ma2016,Chung2019,Darve2009,Izvekov2013}, which usually take the following form:
\begin{equation}\label{eq:basicGLE}
\ddot{x}_t = F_e(x_t) - \int_0^t K(t-s) \dot x_s \dd s + \zeta_t    
\end{equation}
where $x_t\in\R^d$ is the position at time $t$ of the variables, $F_e:\R^d\rightarrow\R^d$ is an effective force, $K:\R_+ \rightarrow \R^{d\times d}$ is a time-dependent linear friction kernel and $\zeta$ is a Gaussian noise. In many cases \cite{Chorin2000,Zwanzig2001,Darve2009,Hijon2009,Li2017,Izvekov2013,Lee2019}, $F_e=-\nabla W\e{pmf}$ where $W\e{pmf}$ is the potential of mean force associated to the coarse variables. When $K$ is a Dirac mass at $0$ and $\zeta$ is a white noise this is the standard Langevin equation. The form \eqref{eq:basicGLE} is heuristically motivated by the so-called Mori-Zwanzig formalism \cite{Zwanzig1973,Mori1965,Zwanzig2001,Mori1965a}, but there are few cases where a rigorous derivation has been established. There is an important literature on linear (harmonic) cases  \cite{Mori1965,Mori1965a,Ciccotti1981,Meyer2020,Doerries2021}. The non-harmonic case, with a non-linear effective force given by the potential of mean force, is discussed in Refs.~\citenum{Lange2006,KinjoHyodo}, but it is observed in Ref.~\citenum{Glatzel2021a} that these derivations rely on some implicit and uncontrolled approximations, and it is argued that in general  it is in fact not possible to get, within the Mori-Zwanzig formalism, an equation of the form \eqref{eq:basicGLE} with a non-linear effective force but a friction term which is linear in the velocity. Our first contribution is a nuance to the statement of \cite{Glatzel2021a}, namely: in general, it is possible to rigorously derive a GLE similar to \eqref{eq:basicGLE} in a general non-linear coupled case, with a non-linear force and a friction which is linear in the velocity, if the kernel $K$ is allowed to depend on the position, \ie a GLE of the form
\begin{equation}\label{eq:resultGLE}
    \ddot{x}_t = F_e(x_t) - \int_0^t K(t-s, x_s) \dot x_s \dd s + \zeta_t   .
\end{equation}
Moreover, in accordance with the standard assumption and in contrast to \cite{Glatzel2021a}, in this case the memory kernel and the covariance of the noise are related by a form of fluctuation-dissipation relation (see sec.~\ref{sec:noise} below for the definition), adapted to the position-dependent framework. To be clear, our framework is restricted to the case where the initial all-atom equation is an equilibrium  Hamiltonian dynamics.

As we will see, along the formal derivation, the kernel $K$ is proven to satisfy some integro-differential equations of Volterra type. As a natural continuation, our second contribution is the implementation of a numerical scheme based on these equations to estimate the kernel and parametrize GLE models from all-atom simulations,  thus adaptating  the inverse Volterra method~\cite{Berne1970,Berkowitz1981,Daldrop2018,Klippenstein2021,Li2017,Ayaz2021} to the position-dependent kernel settings.

\section{Derivation of generalized Langevin equations}
\label{sec:deriv-GLE}

\subsection{Preliminary considerations}

The purpose of the Mori-Zwanzig approach is to extract the evolution in time of an observable from the dynamics of a microscopic state. More explicitly, it relies on the following elements:
\begin{itemize}
    \item A first order differential equation for  a microscopic state $\vect{X}(t)$, given by
    \begin{equation}
  \label{eq:operatorEvolution}
  \partial_t \phi \left( \vect{X}(t)\right)= \mathcal{L} \phi\left( \vect{X}(t)\right) 
\end{equation}
for all \revision{functions} $\phi$ where $\mathcal{L}$ denotes the evolution operator. For the sake of clarity, and since this is the main case of interest,  we will only consider later the deterministic case where $\mathcal L$ is the Liouville operator for Hamiltonian dynamics, in particular $\vect{X}(t)=(\vect{q}(t),\vect{p}(t))\in\R^{2D}$ with $D$ the position space dimension.  Extensions to  Markov operators for stochastic dynamics can be considered, following \cite{Ma2016,Zhu2021} (although, when working at the level of the Kolmogorov equations for stochastic processes, one should be careful whether the goal is to describe the evolution of $\phi(\vect{X}(t))$ or of its expectation with respect to the initial random noise).
\item \revision{A function $\mathcal O:\R^{2D}\rightarrow \R^d$,  so that we are interested in writing a closed effective equation for the observable $\mathcal O(\vect{X}(t))$. For instance, $\mathcal O$ can be a collective variable, or a collective variable and its velocity (in which case  $d=2n$ where $n$ is the dimension of the collective variable, see sec.~\ref{sec:choice-projector}).  }
\item A reference probability density $\rho$ on $\R^{2D}$. For processes at equilibrium, it is simply chosen as the equilibrium Gibbs measure, but other measures are possible~\cite{Givon2005,Meyer2017}. Given $\rho$, we consider on $L^2(\rho,\R^d)=\{\revision{\phi}:\R^{2D}\rightarrow \R^d, \int|\revision{\phi}|^2\dd \rho <\infty\}$   the scalar product
\begin{equation} 
  \label{eq:scalarproduct}
  \sprod{\revision{\phi}}{\revision{\psi}} = \int_{\R^{2D}} \revision{\phi}(\vect{X})\cdot \revision{\psi}(\vect{X})\rho(\vect{X}) \dd \vect{X} .
\end{equation}
\item A set  $\mathcal{E}=\{e_k,\ k\in\mathcal I\}$ of linearly independent functions from $\R^d$ to $\R^d$, for some index set $\mathcal I$.  
\end{itemize}
From the functions $e_k:\R^d\rightarrow\R^d$, it is convenient to introduce the functions $E_k = e_k \circ \mathcal{O}$ from $\R^{2D}$ to $\R^d$\revision{, where $\circ$ is the composition, i.e. $(e_k \circ \mathcal{O}) (\vect{X}) = e_k (\mathcal{O}(\vect{X}))$}.

 Given the observable $\mathcal O$, the reference measure $\rho$ and the family $\mathcal{E}$, we introduce $\mathcal P_{\mathcal{E}}$ the orthogonal projection (in $L^2(\rho)$) on the closure of the space spanned by the set $\{E_k,\ k\in\mathcal I\}$.  For $\phi:\R^ {2D}\rightarrow \R^d$, this orthogonal projection  is formally given by 
\begin{equation}\label{eq:ortho_proj_detailed}
    \mathcal{P}_{\mathcal{E}}\phi = \sum_{k\in \mathcal I} \left(\sum_{k'\in \mathcal I}  G^{-1}_{k,k'} \sprod{E_{k'}}{\phi}\right) E_k
\end{equation}
where $G^{-1}$ the inverse of the Gram matrix of the basis, namely $G_{k,k'} = \sprod{E_{k}}{E_{k'}}$. For instance, when $\mathcal{E}$ is reduced to a \revision{set} of linear functions, this projector is also known as the \textit{Mori} projector~\cite{Grabert2006}. By contrast, when $\mathcal{E}$  is chosen as an Hilbert basis of $L^2(\R^d)$ \revision{(i.e. a sequence of functions whose linear combinations can approximate arbitrarily well any function in $L^2(\R^d)$)} , then this projector is known as the \textit{Zwanzig} projector\cite{Izvekov2013,Glatzel2021a}. Lastly for $\mathcal{E}$  a finite family of function,  $\mathcal{P}_{\mathcal{E}}$ is a \textit{finite-rank} projector~\cite{Chorin2002,Givon2005}. \revision{We refer to App.~\ref{app:projection_op} for details on projection operators.}

\subsection{The general Mori-Zwanzig derivation}
\label{sec:mori-zwanz-deriv}

For $\vect{X} \in\R^{2D}$, we denote by  $\vect{X}(t)$ the flow associated to \eqref{eq:operatorEvolution} with initial condition $\vect{X}$.
 It holds
\begin{equation*}
  %\label{eq:obs_evol}
  \partial_t \mathcal O\po \vect{X}(t)\pf  = \mathcal{L} \mathcal{O}\po \vect{X}(t)\pf  = e^{t\mathcal{L}}\mathcal{L} \mathcal{O}\po\vect{X}\pf = e^{t\mathcal{L}}\mathcal{P}_{\mathcal{E}}\mathcal{L} \mathcal{O}\po \vect{X}\pf+e^{t\mathcal{L}}(1-\mathcal{P}_{\mathcal{E}})\mathcal{L} \mathcal{O}\po \vect{X}\pf.
\end{equation*}
Using the following Duhamel-Dyson formula:
\begin{equation*}
  %\label{eq:operator_iden}
  e^{t\mathcal{L}} = e^{t(1-\mathcal{P}_{\mathcal{E}})\mathcal{L}} + \int_0^t  e^{(t-s)\mathcal{L}} \mathcal{P}_{\mathcal{E}} \mathcal{L} e^{s(1-\mathcal{P}_{\mathcal{E}})\mathcal{L}}\dd s
\end{equation*}
and introducing the notation
\begin{equation*}
  %\label{eq:noise_gle}
  \xi_t(\vect{X}) = e^{t(1-\mathcal{P}_{\mathcal{E}})\mathcal{L}}(1-\mathcal{P}_{\mathcal{E}})\mathcal{L} \mathcal{O}(\vect{X})\,,
\end{equation*}
this first equation can be rewritten as
\begin{align}
  %\label{eq:GLE_general_form_noise}
  \partial_t \mathcal O\po \vect{X}(t)\pf & =  e^{t\mathcal{L}}\mathcal{P}_{\mathcal{E}}\mathcal{L} \mathcal{O}\po \vect{X}\pf+\int_0^t  e^{(t-s)\mathcal{L}} \mathcal{P}_{\mathcal{E}} \mathcal{L} \xi_s \po \vect{X}\pf\dd s+
  \xi_t\po \vect{X}\pf\nonumber\\
  & =  \mathcal{P}_{\mathcal{E}}\mathcal{L} \mathcal{O}\po \vect{X}(t)\pf+\int_0^t   \mathcal{P}_{\mathcal{E}} \mathcal{L} \xi_s \po \vect{X}(t-s)\pf\dd s+
  \xi_t\po \vect{X}\pf.
    \label{eq:GLE_general_form_noise_time}
\end{align}
From Eq.~\eqref{eq:ortho_proj_detailed}, we introduce the coefficients $f_k$ and for $s\geq 0$ the time-dependent coefficients $g_k(s)$ such that
\begin{equation}  \label{eq:coeff_fg}
 \mathcal{P}_{\mathcal{E}}\mathcal{L} \mathcal O  = \sum_{k\in \mathcal I} f_k E_k,  \qquad
 \mathcal{P}_{\mathcal{E}} \mathcal{L} \xi_s =  \sum_{k\in \mathcal I} g_k(s)  E_k.\end{equation}
The projected evolution equation is then, writing $\mathcal O(t)=\mathcal O(\vect{X}(t))$,
\begin{eqnarray}
%  \label{eq:gle_coeffs}
   \Dp{\mathcal{O}(t)}{t}&=& \sum_{k}  f_{k} e_k(\mathcal O(t)) +\int_0^t  \sum_{k} g_{k}(s) e_k(\mathcal O(t-s)) \dd s + \xi_t(\vect{X})\nonumber\\
\label{eq:GLEfgs}
    &:=& f\po \mathcal O(t)\pf + \int_0^t g\po s, \mathcal O(t-s)\pf\dd s + \xi_t\po \vect{X}\pf,
\end{eqnarray}
with $f=\sum_{k} f_k e_k$ and $g(s,\cdot)=\sum_{k}g_k(s)e_k$ for $s\geqslant 0$. This is the generic form of a GLE where on the right hand side, the first term is known as the mean force term, the second term is related to the memory and the last term $ \xi_t\po \vect{X}\pf$, which is interpreted as a noise, follows the orthogonal dynamics~\cite{Givon2005}
\begin{equation}
\label{eq:ortho_dynamics}
    \partial_t \xi_t(\vect{X}) = (1-\mathcal{P}_{\mathcal{E}})\mathcal{L} \xi_t(\vect{X})  \quad \text{with} \quad \xi_0(\vect{X}) = (1-\mathcal{P}_{\mathcal{E}})\mathcal{L}\mathcal{O}(\vect{X}).
\end{equation}
Notice that \eqref{eq:GLEfgs} is an exact equality. The approximation comes afterward when, in order to withdraw the orthogonal degrees of freedom and close the equation without solving the orthogonal dynamics, $\xi_t(\vect{X})$ is replaced by a random noise $\zeta_t$ independent from $\mathcal O(0)$.

 The  key property of $\xi_t(\vect{X})$, under which relies this approximation, is  that it is by design uncorrelated to $e_k(\mathcal O(0))$ for all $k\in\mathcal I$. Indeed, assuming that $\vect{X}$ is a random variable distributed according to $\rho$, we have for all $k \in \mathcal{I}$
\begin{equation}
\label{eq:orthogonality_noise}
\mathbb E \po e_k(\mathcal O(0))\cdot \xi_t(\vect{X})\pf =  \langle E_k,\xi_t\rangle = 0,
\end{equation}
and if for instance  all constant functions are in the span of $\{e_k,k\in\mathcal I\}$, we obtain that the average of $\xi_t(\vect{X})$ with respect to $\rho$ is zero.

The validity in some asymptotic regimes of the approximation  obtained by  replacing $\xi_t(\vect{X})$ by an independent  noise can be shown in some cases, see e.g. \cite{Kupferman2004,TuckermanBook,Leimkuhler2019}. Gaussian processes are then naturally motivated by the central limit theorem, although real applications don't always have particular reasons to fall in this regime \cite{Shin2010,Carof2014,Jung2021}. We won't discuss further this point from a theoretical point of view.

Taking the scalar product of \eqref{eq:GLEfgs} with $E_{\ell}$ for  $\ell\in\mathcal I$, the orthogonality condition~\eqref{eq:orthogonality_noise} yields~\cite{Lee2015}
\begin{equation}
  \label{eq:volterra}
   \left\langle E_\ell(0),\Dp{\mathcal{O}(t)}{t}\right\rangle = \sum_{k}  f_{k} \sprod{E_\ell(0)}{E_{k}(t)} +\int_0^t \sum_{k} g_{k}(s) \sprod{E_\ell(0)}{E_{k}(t-s)} \dd s ,
\end{equation}
where $E_k(t)=e^{t\mathcal L}E_k$. Therefore, we obtain a set of Volterra integral equations of the first kind. 
When the set $\mathcal{I}$ is finite, they can be numerically inverted using a time discretization of the integral as illustrated in sec.~\ref{sec:volt-based-estim}. A set of Volterra equations of the second kind can also be derived as explained in the appendix~\ref{app:proj_corr}.
 
\subsection{Hamiltonian systems}
\label{sec:choice-projector}

Let us now focus on Hamiltonian systems as it is the main case of interest. We consider a system of  particles with position $\vect{q}$ and momentum $\vect{p}$ in $\R^D$. The  Hamiltonian is $\vect{p} \cdot \bm{M}^{-1}\cdot \vect{p} /2 + V(\vect{q})$, with $V$ the potential energy and $\bm{M}$ the mass matrix. The evolution operator~\eqref{eq:operatorEvolution} is then given \revision{for any function $\phi (\vect{q},\vect{p})$ of position and momentums} by
\begin{equation*}
 % \label{eq:evolutionOpe_Hamiltonian}
  \mathcal{L} \phi (\vect{q},\vect{p}) = \vect{p}\cdot\bm{M}^{-1}\cdot \nabla_{\vect{q}}  \phi (\vect{q},\vect{p}) - \nabla_{\vect{q}} V(\vect{q}) \cdot \nabla_{\vect{p}}\phi (\vect{q},\vect{p})
\end{equation*}
and the reference measure is given by the equilibrium Gibbs measure 
\begin{equation*}
    %\label{eq:HamiltonianGibbsMEasures}
    \rho_{eq}(\vect{q},\vect{p}) = e^{ -\beta V(\vect{q}) -\beta \vect{p} \cdot \bm{M}^{-1}\cdot \vect{p}/2}/ Z
\end{equation*}
with $Z$ a normalisation factor and $\beta = 1/k\e{B}T$ the inverse temperature of the system. Integrating by parts, it is well-known that $\mathcal L$ is skew-symmetric in $L^2(\rho_{eq})$, namely
\begin{equation}
\label{eq:antihermitianity}
    \langle \mathcal L\phi,\psi\rangle = -\langle \phi,\mathcal L\psi\rangle
\end{equation}
for all \revision{functions} $\phi,\psi$. 
In particular, applying this with $\psi=1$ so that $\mathcal L\psi=0$, we get that $\rho_{eq}$ is left invariant by the Hamiltonian dynamics.

In view of the general framework described in the previous section, let us now present the specific choices made in the present work, motivated by the specific form of GLE we want to derive. First, we work with a particular choice of observable. Decomposing the microscopic state in its position and momentum coordinates, namely $\vect{X}=(\vect{q},\vect{p})$, we consider a first observable $\mathcal O_x(\vect{X})=\varphi(\vect{q})$ for some \revision{collective variable} $\varphi:\R^{D} \rightarrow \R^n$, and a second observable $\mathcal O_v(\vect{X})= \mathcal L\mathcal O_x(\vect{X}) = \vect{p}\cdot\bm{M}^{-1}\cdot\nabla \varphi(\vect{q})$. Then our observable is $\mathcal O(\vect{X})=(\mathcal O_x(\vect{X}),\mathcal O_v(\vect{X}))$ \revision{of dimension $d=2n$}. We call $\mathcal O_x$ the position observable and $\mathcal O_v$ the velocity observable. Notice that, by design, 
\begin{equation}
    \label{eq:velocity_obs}
    \partial_t \mathcal O_x(\vect{X}(t))=\mathcal O_v(\vect{X}(t))= \vect{p}(t)\cdot\bm{M}^{-1}\cdot\nabla  \varphi(\vect{q}(t)).
\end{equation}
From this choice of observable, the GLE~\eqref{eq:GLEfgs} is now written as
\begin{equation}\label{eq:GLEO1O2}
   \partial_t   \begin{pmatrix}
    \mathcal O_x(t)\\
    \mathcal O_v(t)
    \end{pmatrix}
    = 
    \begin{pmatrix}
    f^x(\mathcal O_x(t),\mathcal O_v(t))\\
    f^v(\mathcal O_x(t),\mathcal O_v(t))
    \end{pmatrix} +     \int_0^t  \begin{pmatrix}
    g^x(s,\mathcal O_x(t-s),\mathcal O_v(t-s))\\
    g^v(s,\mathcal O_x(t-s),\mathcal O_v(t-s))
    \end{pmatrix} \dd s +  \begin{pmatrix}
    \xi^x_t\po \vect{X}\pf\\
    \xi^v_t\po \vect{X}\pf
    \end{pmatrix},
\end{equation}
where $f^x$ and $f^v$ are the mean force terms for position and velocity respectively, $g^x$ and $g^v$ the memory parts and $\xi^x$ and $\xi^v$ the noise on  position and velocity respectively. It remains to choose a set of functions to perform the projection. Given $\{h_k, k\in\mathcal{J}\}$  a family of functions from $\R^n$ to $\R^n$, we consider the family of basis functions from $\R^{2n}$ to $\R^{2n}$ given by
\begin{equation*}
    \mathcal{E} = \left\{a : \begin{pmatrix} o_1\\ o_2\end{pmatrix}
\mapsto \begin{pmatrix} o_2\\ 0\end{pmatrix} \right\}\cup \left\{ b_k : \begin{pmatrix} o_1\\ o_2\end{pmatrix}
\mapsto \begin{pmatrix} 0\\ h_k(o_1)\end{pmatrix} \right\}_{k\in\mathcal{J}}
\cup 
\left\{c_k:\begin{pmatrix} o_1\\ o_2\end{pmatrix}
\mapsto \begin{pmatrix} 0\\ \nabla h_k(o_1)  o_2\end{pmatrix} \right\}_{k\in\mathcal{J}},
\end{equation*} 
where  $\nabla h_k$  stands for the Jacobian matrix of $h_k$.

Let us now comment this particular choice. Our family is divided into three sets of functions $  \mathcal{E} =\{a\} \cup \{b_k,{k\in\mathcal{J}}\} \cup \{c_k,{k\in\mathcal{J}}\}  $. The singleton $\{a\}$ is related to the fact $\partial_t \mathcal O_x(t)=\mathcal O_v(t)$. The second set $\{b_k,{k\in\mathcal{J}}\} $ is introduced to model the position dependency of the mean force term for the velocity. The last set $\{c_k,{k\in\mathcal{J}}\}$ is required to obtain the desired form of the memory part (i.e. linear in velocity).  Notice  that the functions $c_k$ are related to the functions $b_k$ by the fact that $\partial_t h_k(\mathcal{O}_x(t)) = \nabla h_k(\mathcal{O}_x(t))  \mathcal{O}_v(t)$ hence $ \partial_t b_k(\mathcal{O}_x(t) , \mathcal{O}_v(t)) = c_k(\mathcal{O}_x(t) , \mathcal{O}_v(t))$.

As in the general framework, we write $A=a\circ \mathcal O $, $B_k= b_k\circ\mathcal O$ and $C_k=c_k\circ\mathcal O$.

The three parts of $\mathcal E$ correspond to subspaces of $L^2(\rho)$ which are orthogonal one with each other. Indeed the scalar products $\sprod{A}{B_k}$ and $\sprod{A}{C_k}$ are both zero, and
\begin{eqnarray*}
    \sprod{B_k}{C_{k'}} 
    &=&\int_{\R^{D}} h_k(\varphi(\vect{q}))\cdot \nabla h_{k'}(\varphi(\vect{q})) \left( \int_{\R^{D}} \vect{p}\rho_{eq}(\vect{q},\vect{p})  \dd \vect{p}\right)  \cdot \nabla  \varphi(\vect{q})     \dd \vect{q}\ =\ 0\,,
\end{eqnarray*}
where we used \eqref{eq:velocity_obs} and that the average of the momentum under $\rho_{eq}$ is zero. As a consequence, in the representation \eqref{eq:ortho_proj_detailed} of the orthogonal projection, the Gram matrix and its inverse have three diagonal blocks, corresponding to the three parts of $\mathcal E$. 
We introduce $G_b$ (resp. $G_c$) the Gram matrix of the family $\{b_k,{k\in\mathcal{J}}\}$ (resp. $ \{c_k,{k\in\mathcal{J}}\} $).

Finally, we also define for any function $f$ and $g$ of the observable, the scalar product
\begin{equation*}
    \sprod{f}{g}_\varphi =\sprod{f\circ \mathcal O_x}{g\circ \mathcal O_x} = \int_{\R^n}  f(z)^\T\cdot g(z) \rho_{\varphi}(z) \dd z
\end{equation*}
where $\rho_{\varphi}:\R^{n}\rightarrow \R_+ $ is the marginal density of $\rho_{eq}(\vect{q},\vect{p})$, namely is such that
\[\int_{\R^{2D}} g\po \varphi(\vect{q},\vect{p})\pf  \rho_{eq}(\vect{q},\vect{p}) \dd \vect{q}\dd \vect{p}  = \int_{\R^n} g(z) \rho_{\varphi}(z)\dd z  \]
for all function $g$ of the observable, see also appendix~\ref{app:coaire} for a definition using the coarea formula. When $f$ and $g$ are matrix-valued function, the scalar product is defined as
\begin{equation*}
    \sprod{f}{g}_\varphi =  \sum_{i,j} \int_{\R^n}  f(z)_{i,j}g(z)_{i,j} \rho_{\varphi}(z) \dd z.
\end{equation*}

\subsection{Continuing the derivation}
\label{sec:cont-deriv}
Equipped with our choice of observables and basis, we can now make explicit the form of the mean force and the memory kernel in \eqref{eq:GLEfgs}, namely the coefficients $f_k$ and $g_k(s,\cdot)$ in \eqref{eq:coeff_fg}.
The projection of the position evolution equation follows trivially from Eq.\eqref{eq:velocity_obs} and the presence of the singleton $\{a\}$ in our choice of basis. Using the notations of \eqref{eq:GLEO1O2}, we get $ f^x(o_1,o_2)= o_2 $ and $g^x(s,\cdot) = \xi^x_s =0$ for all $s\geqslant 0$.

In the following we focus on the second line of \eqref{eq:GLEO1O2}, namely 
 the projection of the velocity evolution equation. From \eqref{eq:coeff_fg} and the explicit expression of the projector~\eqref{eq:ortho_proj_detailed}, the question is thus to compute the scalar products of $\mathcal L \mathcal O_v$ and of $\mathcal L \xi_s^v = \mathcal L e^{s(1-\mathcal P_{\mathcal E})\mathcal L}\mathcal L\mathcal O_v$ with $B_k$ and $C_k$ (here and in the following, with a slight abuse of notations, we denote by $B_k$ and $C_k$ the second $n$-dimensional component of $B_k$ and $C_k$).

For all $k\in\mathcal J$,
\begin{multline}
    \label{eq:mean_force_K}
        \sprod{C_k }{\mathcal{L} \mathcal{O}_v} 
        =  \int_{\R^{2D}} [\nabla h_k\po \varphi(\vect{q})   \pf (\vect{p}\cdot\bm{M}^{-1}\cdot\nabla\varphi(\vect{q}))] \\\cdot [-\nabla V(\vect{q})\cdot\nabla \varphi(\vect q) + \vect{p}\cdot \bm{M}^{-1}\cdot\nabla^2 \varphi(\vect{q}) \vect{p} ]
           \rho_{eq}(\vect{q},\vect{p}) \dd \vect{q} \dd \vect{p}   
        \ = \ 0,
\end{multline}
where we used that the integrand is odd in $\vect{p}$ and $\rho_{eq}$ is invariant by reflection of the momentum.  The mean force term $f^v$ in~\eqref{eq:GLEO1O2} is then
\begin{equation}
    \label{eq:mean_force_term}
    f^v(o_1,o_2)  =  \sum_{k\in \mathcal{J}} \left(\sum_{k'\in \mathcal{J}}  {({G_b}^{-1})}_{k,k'} \sprod{B_k' }{\mathcal{L} \mathcal{O}_v}\right) h_k\left(o_1\right) =:f_b(o_1).
\end{equation}
In general, this mean force $f_b$ is non-linear (if $\{h_k,k\in\mathcal J\}$ contains non-linear functions on $\R^n$). It can be build from the potential of mean force $W_{\mathrm{pmf}}$ and an effective mass matrix  $M_\varphi$. The potential of mean force  is the function $W_{\mathrm{pmf}} = - \frac1\beta \ln \rho_{\varphi}$. The effective mass matrix is \revision{defined} as the conditional expectation
\begin{align}
  \label{eq:eff_mass_matrix}
 M_\varphi(z)^{-1} &= \beta\,\mathbb E \po \mathcal{O}_v(\vect{q},\vect{p}) \mathcal{O}_v(\vect{q},\vect{p})^\T \ |\ \varphi(\vect{q})=z\pf \\
 &= \mathbb E \po \nabla \varphi (\vect{q}) \cdot \bm{M}^{-1}\cdot \nabla \varphi (\vect{q})^\T \ |\ \varphi(\vect{q})=z\pf \nonumber,
\end{align}
where the second equality follows from $\int_{\R^{D}} \vect{p}  \vect{p}^\T   \rho_{eq}(\vect{q},\vect{p}) \dd  \vect{p}    =\rho_{eq}(\vect{q})  \bm{M}/ \beta$. \revision{Here, $M_\varphi$ is interpreted as an effective mass in view of the equipartition theorem.}
It is then detailed in the appendix~\ref{app:meanforce} that
\begin{equation}
        \label{eq:BkLO2}
    \sprod{B_k }{\mathcal{L} \mathcal{O}_v} =   \sprod{B_k }{\Psi \circ \mathcal O_x} = \sprod{h_k }{\Psi }_{\varphi}
\end{equation}
with $\Psi:\R^n\rightarrow \R^n$ 
\begin{equation}
\label{eq:Psi}
\Psi(z) = \f{1}{\beta} \po \nabla\cdot  M_\varphi(z)^{-1} +  M_\varphi(z)^{-1}\cdot \nabla \ln \rho_{\varphi}(z)\pf .
\end{equation}
If $\Psi$ is in the space spanned by $\{h_k,k\in\mathcal J\}$, then \eqref{eq:BkLO2} implies that $f_b = \Psi$. Otherwise, $f_b$ is a projection of $\Psi$ on this space (this is typically the case in practice since a finite set of basis functions is used).

When the effective mass matrix is independent of the position, we get
\begin{equation}
\label{eq:PsiPMF}
\Psi(z) = \frac{1}{\beta} M_\varphi^{-1}\cdot \nabla  \ln\rho_{\varphi}(z) = - \frac{1}{\beta}  M_\varphi^{-1}\cdot\nabla W_{\mathrm{pmf}}(z) .
\end{equation} 
For instance, this holds if $n=1$ and the reaction coordinate is a distance between two atoms, or for linear reaction coordinates  (taking $\varphi_i(\vect{q})=e_i\cdot \vect{q}$ with $e_1,\dots,e_n$ an orthonormal basis of the range of the reaction coordinate). 

We now turn to the projection of $\mathcal{L} \xi_s^v$. Using the skew-symmetry of $\mathcal{L}$, for all $k\in\mathcal J$, 
\begin{equation*}
    %\label{eq:zero_term_memory}
    \sprod{B_k}{\mathcal{L} \xi_s^v } = -  \sprod{\mathcal{L} B_k}{ \xi_s^v } =   -\sprod{ C_k }{ \xi_s^v } = 0
\end{equation*}
since the noise is orthogonal to any function in $\mathcal E$. 
Therefore the memory part of the equation is (using again the skew-symmetry of $\mathcal L$)
\begin{align}
    \label{eq:memory_term}
    g^v_s(o_1,o_2) &= - \sum_{k\in \mathcal I} \left(\sum_{k'\in \mathcal I}  ({G_c}^{-1})_{k,k'}  \sprod{\mathcal{L} C_k' }{ \xi_s^v}\right) \nabla h_{k} (o_1) o_2  \ =:\  -K_b(s,o_1)  o_2. 
\end{align}
The value of the scalar product $\sprod{\mathcal{L}C_k }{ \xi_s^v}$ is discussed in sec.~\ref{sec:noise}. To follow the usual convention, we refer to $K_b(s,o_1)$  as the \textit{memory kernel}.

Combining together all these elements, we get 
\begin{equation}
    \label{eq:GLE_pos_vel}
    \begin{cases}
    \partial_t \mathcal{O}_x(t) &=  \mathcal{O}_v (t) \\
      \partial_t \mathcal{O}_v (t)  & = f_b(\mathcal{O}_x(t)) -\int_0^t  K_b(s,\mathcal{O}_x(t-s))  \mathcal{O}_v (t-s)  \dd s + \xi_t^v\po \vect{X}\pf.
    \end{cases}
\end{equation}
Closing this equation by approximating $\xi_t^v(\vect{X})$ by a random noise $\zeta_t$ we get, as announced, a GLE for the position $x_t= \mathcal{O}_x(t)$ of the form:
\[\ddot{x}_t = f_b(x_t) -\int_0^t K_b(s,x_{t-s}) \dot{x}_{t-s}\dd s +\zeta_t\,.\]

Up to now, we have left unspecified the set $\{h_k, k \in \mathcal J \}$. Several possibilities are available. First, for instance, we can choose $\{h_k, k \in \mathcal J \}$ as a basis of the linear and constant functions  (\ie the Mori projection), in which case we retrieve the linear form of the GLE
\begin{equation}
    \label{eq:GLE_mori}
    \ddot{x}_t = -\omega (x_t - y_0)- \int_0^t K_M(s) \dot x_{t-s} \dd s + \zeta_t,
\end{equation}
with $\omega$ and $y_0$ some constants characterizing the force.

Second, an alternative is to take a basis of $L^2(\rho,\R^d)$ (\ie the Zwanzig projection). Assuming position-independent effective mass matrix, the GLE then reads
\begin{equation}
    \label{eq:GLE_zwanzig}
    \ddot{x}_t = -M_\varphi^{-1}\cdot \nabla  W\e{pmf}(x_t)- \int_0^t K_Z(s,x_{t-s}) \dot x_{t-s} \dd s + \zeta_t,
\end{equation}
where the mean force term is now the derivative of the potential of mean force. More generally, the mean force is the conditional expectation $\Psi$.

Third, if the mean force term is known a priori (\eg when the potential of mean force is computed apart), then \revision{we can  choose   the set $\{h_k, k \in \mathcal J \}$ to be reduced to  the singleton $\{\nabla W_{\mathrm{pmf}}\}$}, which yields for position-independent mass matrix
\begin{equation}
    \label{eq:GLE_pmf}
    \ddot{x}_t = -M_\varphi^{-1}\cdot\nabla  W\e{pmf}(x_t)- \int_0^t K\e{eff}(s) \nabla^2 W\e{pmf}(x_{t-s}) \dot x_{t-s} \dd s + \zeta_t. 
\end{equation}
This choice is somehow the minimal set-up to get a mean force which derives from $\rho_{\varphi}$.  In that case, the dependency in position of the memory kernel is completely fixed via the Hessian of the potential of mean force. We recover the initial form~\eqref{eq:basicGLE} when this Hessian is approximately constant over the studied range of position (i.e. close to an harmonic case),  and only in this case.

Notice, that if a gradient form of the mean force is wanted, we can take $h_k = \nabla \ell_k$ where $\{\ell_k,k\in\mathcal J\}$ is an Hilbert basis of  the Sobolev space $\mathcal H^1(\R^n)$. In that case $f_b$ is an Helmholtz projection of the mean force obtained with the Zwanzig projection.

In both cases \eqref{eq:GLE_zwanzig} and \eqref{eq:GLE_pmf}, as required in Ref.~\citenum{Glatzel2021a}, the mean force derives from $\rho_{\varphi}$ and the memory kernel is linear in the velocity (although with a non-linear dependency in the position). Indeed, linear reaction coordinates are considered in Ref.~\citenum{Glatzel2021a}, which means a position independent effective mass matrix, possibly up to a  linear change of variable.

\medskip

To conclude this section, let us highlight a slight subtlety: in the general derivation of sec.~\ref{sec:mori-zwanz-deriv}, we assumed the functions $e_k$ to be linearly independent. In fact from a theoretical point of view this is not important since the only thing that matters in the definition of the projection is the space spanned by $\{E_k,k\in\mathcal I\}$. However, from a practical point of view, this is used when considering the inverse of the Gram matrix in the explicit formula \eqref{eq:ortho_proj_detailed}. Then, with the specific choice of basis functions described in sec.~\ref{sec:choice-projector}, it can be convenient to take $\{h_k,k\in\mathcal{J}\}$ which contains some constant functions (in particular to have a mean-zero noise, see next section), in which case the corresponding $c_k$ is zero (since $\nabla h_k=0$), which means the functions $E_k$, $k\in\mathcal I$, are 
not linearly independent. Again, this is not important for the theoretical derivation, but it means that, in practice, one should be careful when applying \eqref{eq:ortho_proj_detailed} and possibly use the pseudo-inverse of the Gram matrix or reduce the set of functions.

\subsection{Characteristics of the noise}
\label{sec:noise}

Once the mean force and memory kernel coefficients have been estimated, the standard way to close the equation \eqref{eq:GLEfgs} is to replace $\xi_t(\vect{X})$ by a Gaussian process $\zeta_t$ with the same mean and variance. However, this approximation may in fact not be suitable in general \cite{Shin2010,Carof2014,Jung2021}. In any cases, let us compute these characteristics. In the following, we consider the specific case introduced in sec.~\ref{sec:choice-projector} and such we are only interested to the properties of the noise $\xi^v_t$ in the second line of  of \eqref{eq:GLE_pos_vel}.

\subsubsection{Mean value}\label{sec:mean_noise}

Since $\xi_t(\vect{X})$ is replaced in practice by a random process which is independent from $\mathcal O(0)$, a minima, it should be enforced that $\xi_t(\vect{X})$ is not correlated to $e_k(\mathcal O(0))$ for $k\in\mathcal I$. As we saw, this is true if the average of $\xi_t(\vect{X})$ is $0$ under $\rho$. 

The simplest way to enforce this is to assume that all constant functions are in the span of $\{e_k,k\in\mathcal I\}$, in which case the orthogonality condition \eqref{eq:orthogonality_noise} concludes. Since we only consider $\xi^v_t $,  the previous condition is satisfied as soon as all constant functions from $\R^n$ to $\R^n$ are in the  space spanned by $\{h_k, \ k\in\mathcal{J}\}$. In which case we obtain $ \avg{\xi^v_t} = 0 $.

\subsubsection{Variance and 2-FDT}\label{sec:FDT}

For GLEs, a Fluctuation-Dissipation Theorem of second type (2-FDT) is a relation which expresses the covariance function of the noise $\xi_t(\vect{X})$ in terms of the memory kernel $K$. It has been established in very general frameworks for \emph{linear} cases \cite{Zwanzig2001,Zhu2021,Jung2021}. In fact, although \eqref{eq:GLEfgs} may be non-linear in general, it is always based on a linear projection, and we can always write a linear GLE for the vector $\po e_k(\mathcal O(t))\pf_{k\in\mathcal I}$. Hence, a 2-FDT holds for this (possibly infinite-dimensional) equation, which implies that a 2-FDT holds for \eqref{eq:GLEfgs} if for instance the identity function is in the span of $\mathcal E=\{e_k,k\in\mathcal I\}$.

This is true if the identity function on $\R^n$ is in the span of $\{h_k,k\in\mathcal{J}\}$. In fact, without loss of generality (up to a change of basis), in that case we simply assume that $h_0(o_1)=o_1$, in which case the corresponding $c_0$ is exactly the required function.  Then, using the skew-symmetry of $\mathcal L$ and the orthogonality \eqref{eq:orthogonality_noise} of the noise,
\begin{align*}
  \sprod{ C_0}{\mathcal{L} \xi^v_s }& = -  \sprod{\mathcal{L} C_0}{ \xi^v_s } = -\sprod{\mathcal{L} \mathcal O_v}{ \xi_s^v } \\&= -\sprod{(1-\mathcal P_{\mathcal E})\mathcal{L} \mathcal O_v}{ \xi_s^v } = -\sprod{\xi_0^v}{ \xi_s^v }\,.
\end{align*}
The left-hand side is a coefficient involved in the decomposition \eqref{eq:memory_term} of $K$. More precisely, if the basis function has been orthonormalized, this is exactly the time-dependent coefficient of $C_0$

If the ambient space dimension $n$ is larger than $1$, then we need  the full covariance matrix of the noise, i.e. $\sprod{(\xi_0^v)_i}{ (\xi_s^v)_j }$ for $1\leqslant i,j\leqslant n$. The previous computation is still correct if we assume now that all the linear functions on $\R^n$ are in the span of $\{h_k,k\in \mathcal{J}\}$. In that case, denoting by $M_{i,j}$ the matrix with all coefficients $0$ except the coefficient $(i,j)$ equal to $1$, $h_{i,j}:o_1\mapsto M_{i,j} o_1$ and the corresponding $C_{i,j}:\vect{X}\mapsto M_{i,j} \mathcal O_v(\vect{X})$, we get again
\[\sprod{ C_{i,j}}{\mathcal{L} \xi_s } = -\sprod{(\xi_0^v)_i}{ (\xi_s^v)_j }\,. \]
As a conclusion, in this case, we can say that a 2-FDT holds in the sense that the covariance matrix of the noise in \eqref{eq:GLE_pos_vel} is determined by some particular coefficients of the kernel $K$ in \eqref{eq:memory_term}. In practice, this means that the covariance of the noise can be computed through  the Volterra equations \eqref{eq:volterra}.

\subsubsection{Position-dependent average and 2-FDT}
\label{sec:pos_dep_FDT}
The fact that $\xi_t(\vect{X})$ is uncorrelated to $e_k(\mathcal O(0))$ for all $t\geqslant 0$ and $k\in\mathcal I$ gives little information on a possible dependency of the law of $\xi_t(\vect{X})$ with respect to $\mathcal O(t)$. In particular, if we use the covariance function computed in the previous section to approximate $\xi_t(\vect{X})$ by a Gaussian noise $\zeta_t$ with the same characteristics, then the law of $\zeta_t$ does not depend on $\mathcal O(t)$. This approximation may not be relevant, having in mind some alternative ways to derive an effective dynamics \cite{Legoll2010,ZHS}. We discuss in the following some results about the position dependence average and variance of the noise.

Let us introduce the conditional average of the noises $\bar{\xi}_t$ and $\bar{\xi}^{\dot z}_t$ given by
\begin{equation}
\label{eq:cond_expec_noise}
    \bar{\xi}_t(z) =  \mathbb E \po \xi^v_t(\vect{q},\vect{p})  \ |\ \varphi(\vect{q})=z \pf,
\end{equation}
and 
\begin{equation}
    \label{eq:cond_expec_noise_2}
     \bar{\xi}^{\dot z}_t(z) = \mathbb E \po \xi^v_t(\vect{q},\vect{p}) \mathcal O _v (\vect{q},\vect{p})^\T \ |\ \varphi(\vect{q})=z\pf.
\end{equation}
 Eq.~\eqref{eq:orthogonality_noise} induces on $\bar{\xi}_t$ and $ \bar{\xi}^{\dot z}_t$ the set of constraints:
\begin{equation}
\label{eq:orthogonality_noise_recast}
\forall k \in \mathcal{J}, \quad \sprod{h_k  }{\bar{\xi}_t }_\varphi  = 0 \quad \text{and} \quad \sprod{\nabla h_k   }{\bar{\xi}^{\dot z}_t}_\varphi = 0.
\end{equation}

Similarly consider  $\sigma_\varphi(t,z) $ the conditional  variance of the noise given by 
\begin{equation}
\label{eq:cond_variance_noise}
    \sigma_\varphi(t,z) =  \mathbb E \po \xi^v_0(\vect{q},\vect{p}) \xi^v_t(\vect{q},\vect{p})^\T  \ |\ \varphi(\vect{q})=z\pf.
\end{equation}
As detailed in the appendix~\ref{app:fdt}, we can decompose the memory kernel $K_b$ as
\begin{equation}
\label{eq:kernel_FDT}
  K_b(t,o_1)  = K_\sigma(t,o_1)+ K_{\bar\xi}(t,o_1) ,
\end{equation}
where  $ K_\sigma(t,o_1)$ and $ K_{\bar\xi}(t,o_1)$ are defined by
\begin{equation}
\label{eq:kernel_sigma}
    K_\sigma(t,o_1) =  \sum_{k\in \mathcal I} \left(\sum_{k'\in \mathcal I}  ({G_c}^{-1})_{k,k'}  \sprod{\nabla h_{k'}}{ \sigma_\varphi(t, \cdot)}_\varphi\right) \nabla h_{k} (o_1)
\end{equation}
and
\begin{equation}
\label{eq:kernel_noise_avg}
    K_{\bar\xi}(t,o_1) =  \sum_{k\in \mathcal I} \left(\sum_{k'\in \mathcal I}  ({G_c}^{-1})_{k,k'}  \sprod{\nabla h_{k'}  }{f_b\, \bar{\xi}_t- \nabla \cdot \left(\bar{\xi}^{\dot z}_t \rho_\varphi \right) /\rho_\varphi}_\varphi\right) \nabla h_{k} (o_1) ,
\end{equation}
with $f_b$ is the mean force of Eq.\eqref{eq:mean_force_term}. To obtain a position-dependent 2-FDT, namely a pointwise relation between $K_b$ and $\sigma$, $K_{\bar\xi}$ should cancel. This is not the case in general. Indeed in~\eqref{eq:kernel_noise_avg}, the terms multiplying the conditional average of the noise do not necessary belong to the space spanned by our function set and, as a consequence, the conditions~\eqref{eq:orthogonality_noise_recast} cannot be used to prove that this term is zero. This fact is illustrated in sec.~\ref{sec:LJDimer}.

However, $K_{\bar\xi}(t,o_1)$ is indeed zero when the set $\{h_k, k \in \mathcal J \}$ is a basis  of $L^2(\rho,\R^d)$ (\ie in the case of the Zwanzig projection $K_b=K_Z$). From the conditions \eqref{eq:orthogonality_noise_recast}, we get
\begin{equation*}
    \label{eq:zero_mean_noise}
    \forall z \in \R^d, \quad \bar{\xi}_t(z) = 0 \quad \text{and} \quad \bar{\xi}^{\dot z}_t(z) = 0,
\end{equation*}
from which $K_{\bar\xi}(t,o_1) = 0$. We then get a position-dependent 2-FDT from the effective mass matrix and conditionnal variance of the noise,
\begin{equation*}
    K_Z(t,o_1) = \beta M_\varphi(o_1)\cdot \sigma_\varphi(t,o_1)
\end{equation*}
where $K_Z$ is the memory kernel of Eq.~\eqref{eq:GLE_zwanzig}. 

Finally, when the Mori projection is used, we emphasize that $K_{\bar\xi}(t,o_1)$ is also zero and we recover the result of sec.~\ref{sec:FDT}.

\subsection{Other forms of GLE}
\label{sec:others_choice}

The  particular choices of observable and of function set made in sec.~\ref{sec:choice-projector} lead to the GLE~\eqref{eq:GLE_pos_vel}, but other choices are possible. In fact, a priori, although it may be appealing from a physical point of view, there is no reason for the best model of approximation of an effective dynamics to be a position/velocity GLE with a mean force term deriving from the potential of mean force and a friction kernel which is linear in the velocity (which was the motivation of these choices). Here are some other possibilites.

The form of the observable in Ref.~\citenum{Glatzel2021a} is similar to the one in sec.~\ref{sec:choice-projector}, but the  function set is chosen as
\begin{equation}
    \mathcal{E}_{GS} = \left\{a : \begin{pmatrix} o_1\\ o_2\end{pmatrix}
\mapsto \begin{pmatrix} o_2\\ 0\end{pmatrix} \right\}\cup \left\{ f_k : \begin{pmatrix} o_1\\ o_2\end{pmatrix}
\mapsto \begin{pmatrix} 0\\ f_k(o_1,o_2)\end{pmatrix} \right\}_{k\in\mathcal{J}}
\end{equation}
where $(f_k)_{k\in\mathcal{J}}$ is a basis of $L^2(\R^{2d})$. This results in a GLE of the form
\begin{equation}\label{eq:resultGLE_GS}
    \ddot{x}_t = F_e(x_t) - \int_0^t K_{GS}(s, x_{t-s}, \dot x_{t-s}) \dd s + \zeta_t ,
\end{equation}
where the memory is no longer linear in the velocity.

In contrast, we could have chosen a smaller function set containing only the two first families of sec.~\ref{sec:choice-projector}, namely
\begin{equation*}
    \mathcal{E}_{NV} =\{a\} \cup \{b_k,{k\in\mathcal{J}}\}  .
\end{equation*} 
In that case, we obtain a GLE of the form
\begin{equation}
\label{eq:resultGLE_NV}
    \ddot{x}_t = F_e(x_t) - \int_0^t K_{NV}(s, x_{t-s}) \dd s + \zeta_t ,
\end{equation}
where there is no dependence in velocity in the memory term. 

A slightly larger function set is obtained by adding a function linear in velocity
\begin{equation*}
    \mathcal{E}_{H} =\{a\} \cup \{b_k,{k\in\mathcal{J}}\} \cup \left\{ \begin{pmatrix} o_1\\ o_2\end{pmatrix}
\mapsto \begin{pmatrix} 0\\ o_2\end{pmatrix} \right\} .
\end{equation*} 
This leads to the hybrid projector of Ref.~\citenum{Ayaz2022}. In which case the GLE is of the form
\begin{equation}
\label{eq:resultGLE_H}
    \ddot{x}_t = F_e(x_t) - \int_0^t K^x_{H}(s, x_{t-s}) \dd s-\int_0^t K^p_{H}(s) \dot x_{t-s} \dd s + \zeta_t .
\end{equation}

Next, if we only consider the position observable, namely $\mathcal O(\vect{X})=\mathcal O_x(\vect{X})=\varphi(\vect{q})$, and the function set
\begin{equation*}
    \mathcal{E}\e{ov} =  \left\{h_k : o_1 \mapsto h_k(o_1) \right\},
\end{equation*}
this leads to an overdamped form of the GLE
\begin{equation}
\label{eq:resultGLE_overdamped}
    \dot{x}_t = F\e{ov}(x_t) - \int_0^t K\e{ov}(s, x_{t-s}) \dd s + \zeta_t.
\end{equation}
Similarly, one can consider only the velocity.

After that, let us emphasize that, although kinetic or overdamped equations may, again, be appealing from a physical point of view since many physical laws are associated to these familiar structures, there is a priori no reason to be restricted to equations of order $1$ or $2$. For instance, considering an observable $\mathcal O=(\mathcal O_x,\mathcal O_v,\mathcal O_a)$ with $\mathcal O_x(\vect{X})=\varphi(\vect{q})$, $\mathcal O_v = \mathcal L\mathcal O_x$ and  $\mathcal O_a = \mathcal L\mathcal O_v$ naturally yields a third order effective equation for $\mathcal O_x$, with a structure depending again on the choice of the basis functions.

Finally, in this work we considered basis functions which depend a priori of the whole observable, since we are looking for closed effective dynamics. However, when working on a pair of observables, nothing forbids the choice of a projection on a set of functions which depend only of one of the observables, or even which depend on a different observable that the one we are interested to.

\section{Volterra based estimation of the memory kernel}
\label{sec:volt-based-estim}
\subsection{Numerical inversion of the Volterra Integral Equation}
\label{sec:numVolterra}

Given an all-atom simulation $(\vect{X}(t))_{t\in[0,T]}$ (or more generally a set of independent simulations), a practical issue is to parametrize the GLE \eqref{eq:resultGLE}, namely to estimate the coefficients $f_k$ and $g_k(s)$ in \eqref{eq:GLEfgs}. This can be done using the Volterra equations \eqref{eq:volterra}, when the set $\mathcal E$ is finite. Indeed, the averages involved in \eqref{eq:volterra} can be estimated along the trajectory, and then the equations can be numerically inverted using a discretization of the time integral~\cite{Linz1969}. In the following, we use the trapezoidal method of Ref.~\citenum{Linz1969}, we also point out that this trapezoidal rule suffers from oscillatory error~\cite{Jones1961} and requires a smoothing that we implement. The numerical implementation  of the inversion of the Volterra equation used in the following is available at \url{https://github.com/HadrienNU/VolterraBasis} \revision{and its documention is available at \url{https://volterrabasis.readthedocs.io}}. 

The inversion of Volterra equation of the first kind is an ill-conditioned problem~\cite{Lamm2000,Lange2006}. Regularization methods can be used to improve the quality of the algorithm but this falls out of scope of this work and we refer to the rich literature on the subject~\cite{Lamm2000,Baker2000,Biazar2003,Armand2014}.

Notice that, apart from Volterra-based algorithms, other methods exist to estimate friction kernels \cite{Carof2014,Lesnicki2016,Jung2017,Hanke2021} and could be modified to tackle the position-dependent case. However, here we only consider the former for simplicity and since they are naturally related to the theoretical derivation of the previous section.

\subsection{Lennard-Jones Dimer}
\label{sec:LJDimer}

In this section, we present some  numerical experiments to illustrate this approach (a thorough numerical analysis, including a discussion on the statistical properties of the noise \revision{and how it can be generated}, on the ability of the effective dynamics to reproduce the dynamics of the initial system, or on other inference methods as in Ref.~\citenum{Vroylandt2022a}, is out of the scope of this work). The present analysis is based on the molecular dynamics simulation of a Lennard-Jones dimer similar to the one used in our earlier work~\cite{Vroylandt2022a}. Position dependence for the memory kernel in this system has been emphasized in Ref.\citenum{Straub1990}.

\revision{
This is a 3D system composed of $2$ Lennard-Jones (LJ) particles forming a dimer and $510$ LJ particles constituting the solvent at reduced temperature $\hat{T}=k_BT/\epsilon =1$ and reduced density $\hat \rho = \rho\sigma^3= 1$. The two dimer particles interact via a LJ potential with parameters $\epsilon_d=2$ and $\sigma_d=1$ (in LJ units) and their distance $r$ is constrained for distance above $r=3.0$ using a parabolic wall of strength $\kappa=200$. LJ parameters for dimer-solvent and solvent-solvent interaction are taken as $\epsilon=1$ and $\sigma=1$ (in LJ units).} The size of the cubic simulation box is $8\sigma$, with periodic boundary conditions in all directions. The dynamics is integrated with a time step of $\Delta t_{MD}= 10^{-3}$ (in LJ units) in the NVE ensemble using the LAMMPS simulation package \cite{Plimpton1995}. We run 500 trajectories with length of $2\cdot 10^5 $ timesteps and CV values are extracted every steps.
 The observable $\mathcal O_x(\vect{q})$ is the distance $r$ between the two particles forming the dimer.
 
\begin{figure}[t]
     \centering
    \includegraphics[width=0.5\linewidth]{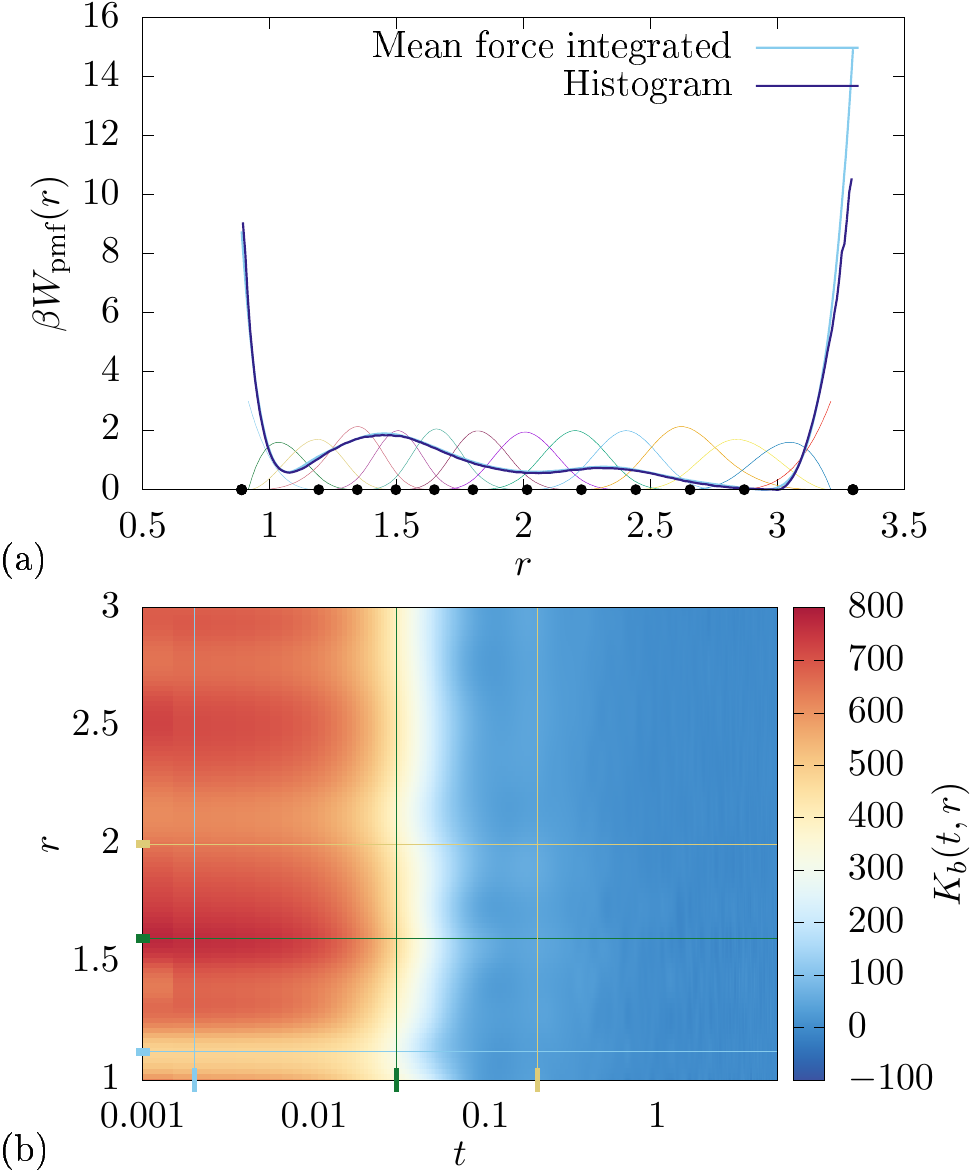}
     \caption{(a) Potential of mean force computed from an histogram compared to the integral of the mean force term. We also represent all the function in the splines basis (thin colored line) with the position of the knots marked with black dots. (b) Memory kernel as a function of the distance and time. Horizontal and vertical lines are the position of the cross-section with the same color represented in Fig.~\ref{fig:cross_section}.}
     \label{fig:2d_kernel}
 \end{figure}

We use as set of functions $\{h_k, k \in \mathcal J \}$ a set of $14$ spline functions of degree 3 with knots non uniformly distributed between $[0.889,3.297]$ as represented in Fig.~\ref{fig:2d_kernel}, the two bounds corresponding to the minimum and maximum value of the distance in the data. More splines function are put on the part of the potential of mean force profile with the most important variation. We refer in the following to this choice of function set as the \textit{splines} GLE.

We also study the \textit{linear} GLE of \eqref{eq:GLE_mori} using as function set a linear function and the \textit{minimal} GLE where we use as function set the the PMF as in Eq.~\eqref{eq:GLE_pmf}.

\begin{figure}[t]
     \centering
     \includegraphics[width=0.5\linewidth]{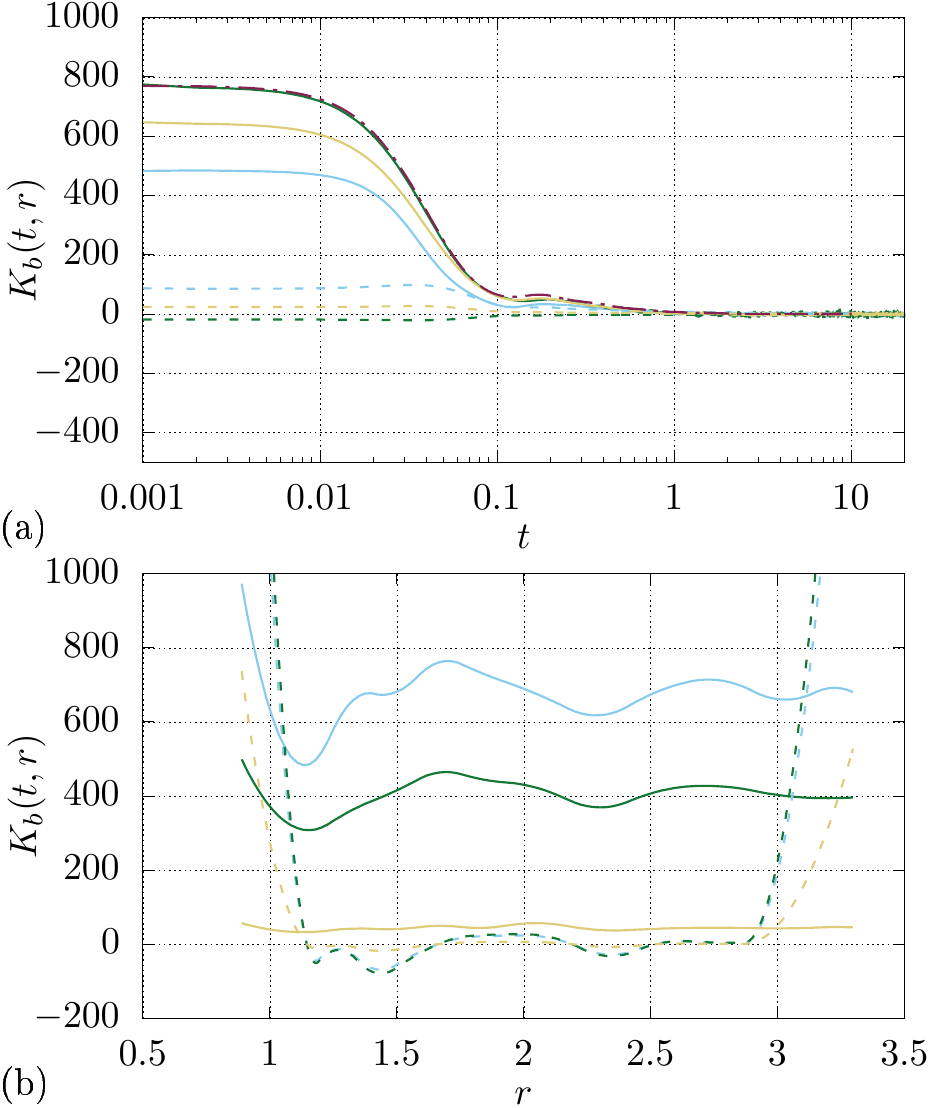}
     \caption{Cross section of the 2D plot of~\ref{fig:2d_kernel}(b) along (a) time and (b) position. Solid line are the cross-section of the \textit{splines} GLE and dashed lines the \textit{minimal} memory kernel computed along the same cross-section. Color matches the lines on Fig.~\ref{fig:2d_kernel}. The dot-dashed violet curve on panel(a) is the memory kernel for the \textit{linear} GLE.}
     \label{fig:cross_section}
 \end{figure}

We represent on Fig.~\ref{fig:2d_kernel}(a) the potential of mean force obtained via an histogram of the distance or via the integration of the mean force computed via the projection of the acceleration on the spline basis, showing a good agreement between both. The position dependent memory kernel is represented as a 2D plot in Fig.~\ref{fig:2d_kernel}(b) and some cross-sections are represented in Fig.~\ref{fig:cross_section} along time (a) and space (b). 

Comparing cross-sections of the \textit{splines} GLE with cross-sections of other GLEs emphasizes the importance of the basis choice. For the \textit{minimal} GLE most of the memory is concentrated at  the boundary (where the potential of mean force is singular) and even if the mean force term is the same than the \textit{splines} GLE, the memory kernel is strongly different from the previous case. Interestingly, using the simple \textit{linear} GLE reconstruct a memory kernel very similar to the one of the \textit{splines} GLE. 
 
Finally, we also study the noise. Indeed, once the mean force and memory kernel have been estimated, the noise can be computed from the \revision{molecular dynamics} trajectory data by inversion of Eq.~\eqref{eq:GLE_pos_vel}. Estimation of correlation function between the noise and any function can also be carried out as detailed in the appendix~\ref{app:proj_corr}.  \revision{In the objective of generating new trajectories using the GLE coarse-grained dynamics, the noise should be modeled, usually as a stationary Gaussian correlated noise~\cite{Lu2005,Maes2013,Schmidt2015}. Validity of such  models could be then interpreted in the light of noise computed from the trajectory data.}

\begin{figure}[t]
     \centering
     \includegraphics[width=0.5\linewidth]{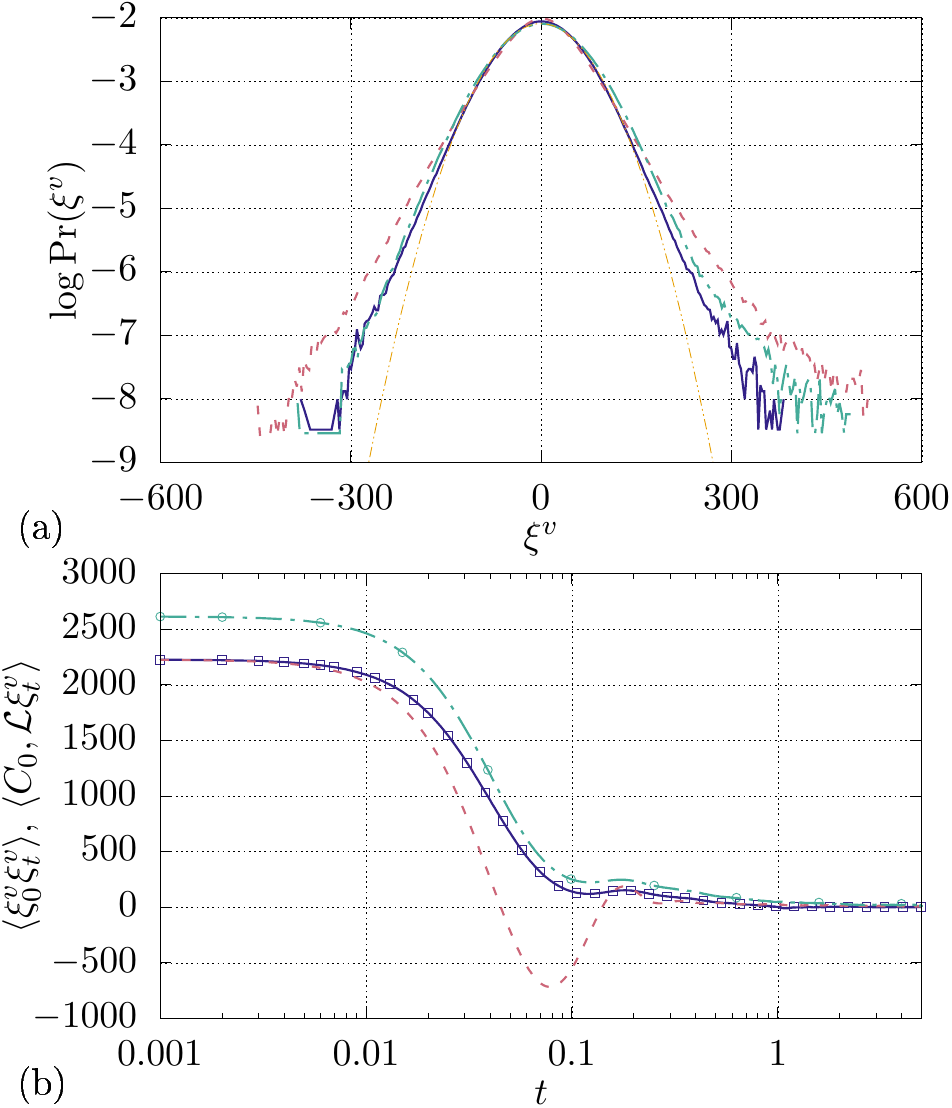}
     \caption{(a) Log of the noise histogram for the \textit{splines} GLE (blue line), the \textit{minimal} GLE (magenta dashed line) and the \textit{linear} GLE (cyan dot dashed line). For comparison, a quadratic curve is fitted to the top of the histogram (dot-dashed yellow thin line). (b)  Auto correlation of the noise for the \textit{splines} GLE (blue line), the \textit{minimal} GLE (magenta dashed line) and the linear GLE (cyan dot dashed line)  compared to the coefficient $\sprod{ C_0}{\mathcal{L} \xi_s }$ of the memory kernel for the spline GLE (blue squares) and the \textit{linear} GLE (cyan circles).}
     \label{fig:noise}
 \end{figure}

Histogram of the value of the noise are plotted on Fig.~\ref{fig:noise}(a) showing similar behavior for all GLE but with non-gaussian tails, a feature that has been noticed on several systems~\cite{Shin2010, Carof2014}. \revision{Notice that with the \textit{splines} GLE, the noise is more Gaussian than with the \textit{linear} GLE or the \textit{minimal} GLE, which illustrates in these two cases the fact that more non-linear features of the dynamics have to be included in the noise.}

On Fig.~\ref{fig:noise}(b) we investigate the validity of the FDT by comparing the noise auto-correlation to the constant part of the kernel. The \textit{splines} and \textit{linear} GLE show a good agreement between noise autocorrelation and corresponding coefficient of the memory kernel.  The memory kernel of the \textit{minimal} GLE does not have a coefficient corresponding to the noise auto-correlation and is therefore not represented.

\begin{figure}[t]
     \centering
     \includegraphics[width=\linewidth]{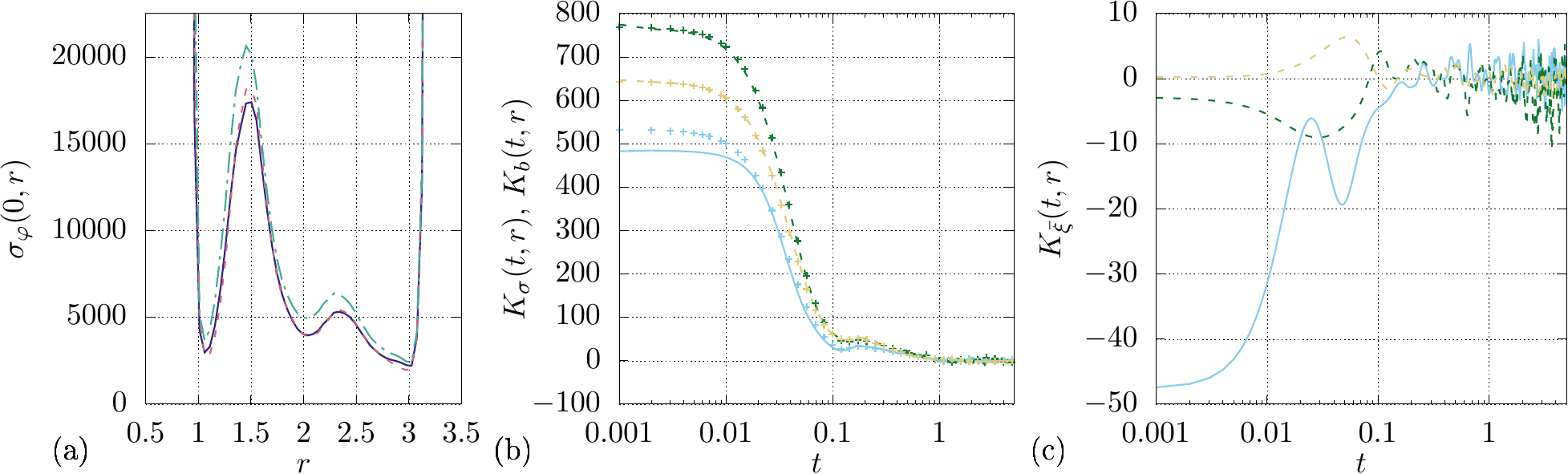}
     \caption{(a) Variance of the noise as a function of the position for the \textit{splines} GLE (blue line), the \textit{minimal} GLE (magenta dashed line) and the \textit{linear} GLE (cyan dot dashed line). (b) Cross section of the memory kernel and $K_\sigma(t,r)$ along time. Solid line are the cross-section of the \textit{splines} memory kernel and cross  are the values of $K_\sigma(t,r)$ computed along the same cross-section. (c) Cross section of $K_{\bar\xi}(t,r)$ along time for the \textit{splines} GLE.   For (b) and (c) color matches the lines on Fig.~\ref{fig:2d_kernel}.}
     \label{fig:noisefdt_pos}
 \end{figure}

Following the discussion of sec.~\ref{sec:pos_dep_FDT}, the position dependence of the conditional expectation of the variance of the noise as a function of the position at $t=0$ is plotted on Fig.~\ref{fig:noisefdt_pos}(a). For each trajectories, we bin the value of the noise depending of the position using $50$ bins. As expected, the variance of the noise shows a strong dependency in position.

Finally, we study the validity of the position-dependent 2-FDT for the \textit{splines} GLE. We represent on Fig.~\ref{fig:noisefdt_pos}(b) the comparison the memory at three different cross-section with the corresponding 2-FDT curve $K_\sigma$ of Eq.~\eqref{eq:kernel_sigma}. The disagreement between the curves is explained by the non-zero value of $K_\xi$ in Eq.~\eqref{eq:kernel_FDT} as illustrated in Fig.~\ref{fig:noisefdt_pos}(c). Indeed, in this case the splines does not form a complete basis of $L^2(\rho,\R^d)$.  

\section*{Conclusion}
\label{sec:conclusion}

We have shown how GLE with a non-linear mean force, a friction kernel linear in the velocity and a Fluctuation-Dissipation Theorem can be rigorously derived within the Mori-Zwanzig formalism, in the Hamiltonian equilibrium settings. However, in contrast to the standard assumption, this requires in general the friction kernel to depend on the position (as confirmed by the numerical experiments). Moreover, we have also emphasized that, even with the position-dependency, this particular form remains arbitrary to some extent, since various other forms may be similarly derived with a suitable choice of observable and of function basis for the projection. The complexity of the model can then be monitored by these choices, a better approximation requiring a larger set of functions, hence of parameters to estimate, leading to a higher sensitivity to the sampling variations. In any cases, the question of obtaining a good approximation cannot be reduced to the choice of the form of the GLE and the estimation of the effective force and the memory kernel: a correct representation of the noise is also necessary, which may require to go beyond space-homogeneous Gaussian processes. In this case, then, this representation cannot be extracted only from the Volterra  equations and the 2-FDT, but should rely on  a suitable statistical analysis.

\section*{Acknowledgments}
We thank Tanja Schilling and Fabian Glatzel for enriching discussions on the subject as well as their useful comments on the manuscript.
We thank Sara Bonella, Michele Casula, Arthur France-Lanord, Ludovic Goudenège, Fabio Pietrucci, Benjamin Rotenberg, Marco Saitta, Mathieu Salanne, and Rodolphe Vuilleumier for fruitful discussions within the MAESTRO collaboration.

\bibliography{GLE.bib}
%\printbibliography

\appendix
\begin{widetext}
\revision{
\section{Projection operators}
\label{app:projection_op}

By definition, the orthogonal projection operator on the closure of the space spanned by the set $\mathcal{E}=\{E_k,\ k\in\mathcal I\}$ is the linear operator characterized by the two following properties: first,  for all $k\in\mathcal I$, $\mathcal P_{\mathcal E} E_k = E_k$. Second, $\mathcal P_{\mathcal E}\phi= 0$ for all functions $\phi$ which are orthogonal to $E_k$ for all $k\in\mathcal I$. The idempotency of the projector, namely $\mathcal P_{\mathcal E}\mathcal P_{\mathcal E}\phi=\mathcal P_{\mathcal E}\phi$, follows from those properties.

Let us check that these properties are indeed satisfied by the operator given by Eq.\eqref{eq:ortho_proj_detailed}. The second one is clear, and for all $k,j\in\mathcal I$,
\[\sum_{k'\in\mathcal I} G_{k,k'}^{-1} \sprod{E_{k'}}{E_j} = (G^{-1} G)_{k,j} = 1_{k=j}\,,\]
from which $\mathcal P_{\mathcal E} E_j = E_j$ for all $j\in\mathcal I$. 

As emphasized before, the projection on various spaces are usually considered. Recall that we consider $E_k = e_k\circ \mathcal O$, i.e. in any cases we only consider functions of the observable $\mathcal O$. In that case we can write $\mathcal P_{\mathcal E} \phi = f_\phi \circ \mathcal O$ where $f_\phi $ is the function of the observable given by
\[f_\phi = \sum_{k\in\mathcal I} \sum_{k'\in\mathcal I} G_{k,k'}^{-1} \sprod{E_{k'}}{E_j} e_k \,.\]
\begin{itemize}
    \item In the case of the projection on a singleton set $\mathcal{E} = \{E_0\}$, the projection becomes
    \begin{equation*}
        f_\phi(z) = \f{\sprod{E_{0}}{\phi}}{\sprod{E_{0}}{E_{0}}} e_0(z)
    \end{equation*}
    Moreover,  when the function $e_0(z)=z$ is simply the identity,   this is the usual linear Mori projection.
\item In the multidimensional case, we can project onto the set of linear functions $e_j:z\mapsto z_j$ returning the $j$ coordinate of $z$, the projection  then reads
 \begin{equation*}
   f_\phi(z) = \sum_{j,j'=1}^d   G^{-1}_{j,j'} \sprod{E_{j'}}{\phi}z_j\,,
\end{equation*}
%where the Gram matrix have entries $G_{j,j'}=\sprod{z_{j}}{z_{j'}}$.
in particular $f_\phi$ is a linear function.
\item We can also consider the projection upon all functions of $z\in \R^d$. The orthogonal projection is then given by the conditional expectation
\begin{equation*}
    f_\phi(z) =  \mathbb E \po \phi(\vect{X})\ |\ \mathcal O(\vect{X})=z\pf. 
\end{equation*}
Informally, this projector can be obtain from Eq.~\eqref{eq:ortho_proj_detailed} using for $\mathcal{E}$ the set of Dirac function $\{\delta(\mathcal O(\vect{X})-z) ,z \in \R^d \}$, but choosing $\{e_k,k\in\mathcal I\} $ as any Hilbert basis of $L^2(\R^d)$ will give the same result. This is the Zwanzig projection, or "non-linear" projection. Note that "non-linear" refers here to $f_\phi$ being a nonl-linear function of $z$ and not to the linearity of the projection operator.
\item Finally when we consider an observable that could be separated into a position observable and a velocity observable, as in the main text sec.~\ref{sec:cont-deriv} or sec.~\ref{sec:others_choice}, we could consider a projection on functions of only the position observable $\varphi(\vect{q})$. For instance, the projection upon all function of the position would become for $x \in \mathbb R^n$ 
\begin{equation*}
     f_\phi(x) =  \mathbb E \po \phi(\vect{q},\vect{p})\ |\ \varphi(\vect{q})=x\pf. 
\end{equation*}
\end{itemize}

}

\section{Coarea formula}
\label{app:coaire}
The projection can be expressed using the coarea formula: for an observable $\phi$: $\Omega \to \mathbb{R}^d$, we have
\begin{equation*}
  \label{eq:coareaformula}
   \int_\Omega  \psi(\vect{X}) \dd \vect{X} = \int_{\mathbb{R}^d}  \int_{\Sigma_z}  \psi(\vect{X})  \omega_{\phi}(\vect{X})\sigma_z(\dd \vect{X}) \dd z
 \end{equation*}
 for any function $\psi$ on $\Omega$, where $\Sigma_z=\{\vect{X}\in\Omega,\ \phi(\vect{X})=z\}$, $\sigma_z$ is the Lebesgue measure on $\Sigma_z$ and $\omega_{\phi}(\vect{X})=1/\sqrt{\det\po (\nabla \phi(\vect{X}))^T\nabla\phi(\vect{X})\pf }$ is the inverse of the $d$-dimensional Jacobian of $\mathcal O$. The measure  $\omega_{\phi}\sigma_z (\dd \vect{X})$ is also denoted by $\delta(\phi(\vect{X})-z) \dd \vect{X}$. Using the notations of sec.~\ref{sec:cont-deriv}, we see that the marginal density $\rho_{\varphi}$ is given by
 \[\rho_{\varphi}(z) = \int_{\R^D} \int_{\Sigma_z}   \omega_{\varphi}(\vect{q}) \rho_{eq}(\vect{q},\vect{p})\dd \vect{p} \sigma_z (\dd \vect{q})\]
where $\Sigma_z$ corresponds to $\varphi(\vect{q})=z$. 

This mean that scalar product between a function $g$ of the observable and a function $F$ can be written as integral over a $d$-dimensional space
\begin{align*}
  \label{eq:integrationF}
   \int_{\R^{2D}}  g(\mathcal{O}(\vect{X}))^\T \cdot F(\vect{X})\rho_{eq}(\vect{X})\dd \vect{X} &= \int_{\mathbb{R}^d}   g(z)^\T \cdot \int_{\Sigma_z}  F(\vect{X})  \omega_{\mathcal O}(\vect{X}) \rho_{eq}(\vect{X})  \sigma_z (\dd \vect{X})\dd z \\
   &=\int_{\mathbb{R}^d}  g(z)^\T \cdot  F_\varphi(z) \rho_{\varphi}(z)  \dd z\\
   &= \sprod{g}{ F_\varphi}_\varphi
 \end{align*}
 having introduced the conditional expectation
\begin{equation*}
     F_\varphi(z) =  \mathbb E \po F(\vect{X})\ |\ \varphi(\vect{q})=z\pf  = \f{1}{\rho_{\varphi}(z)}  \int_{\Sigma_z}  F(\vect{X})  \omega_{\mathcal O}(\vect{X}) \rho_{eq}(\vect{X})  \sigma_z (\dd \vect{X}).
\end{equation*}

\section{Derivation for the mean force term}
\label{app:meanforce}
Using the skew-symmetry~\eqref{eq:antihermitianity} of $\mathcal{L}$, we have for the mean force term
\begin{align*}
    \sprod{B_{k}}{\mathcal{L} \mathcal{O}_v}& = - \sprod{C_{k}}{\mathcal{O}_v} =- \sprod{\nabla h_k(\mathcal{O}_x(t))  \mathcal{O}_v}{\mathcal{O}_v} \\
    &=  -\int_{\R^{D}} \int_{\R^{D}} \mathcal{O}_v(\vect{q},\vect{p})^\T \cdot \nabla h_k(\mathcal{O}_x(\vect{q}))^\T \cdot \mathcal{O}_v(\vect{q},\vect{p}) \rho_{eq}(\vect{q},\vect{p})  \dd \vect{q}\dd\vect{p} \\
    &= -\sum_{j=1}^n  \sum_{i=1}^n \int_{\R^{D}} \int_{\R^{D}} \partial_{z_j} h_k(\mathcal{O}_x(\vect{q}))_{i}  \mathcal{O}_{v,j}(\vect{q},\vect{p}) \mathcal{O}_{v,i}(\vect{q},\vect{p}) \rho_{eq}(\vect{q},\vect{p})  \dd \vect{q}\dd\vect{p}.
\end{align*}
From the coarea formula and the definition~\eqref{eq:eff_mass_matrix} of the effective mass matrix, we obtain
\begin{align*}
     \sprod{B_{k}}{\mathcal{L} \mathcal{O}_v}  
     &= - \int_{\R^n}  \sum_{j=1}^n  \sum_{i=1}^n \partial_{z_j} h_k\po z\pf _i \left( \int_{\Sigma_z} \int_{\R^{D}} \,\mathcal{O}_{v,j}(\vect{q},\vect{p}) \mathcal{O}_{v,i}(\vect{q},\vect{p}) \omega_{\varphi}(\vect{q})\rho_{eq}(\vect{q},\vect{p}) \sigma_z(\dd \vect{q})\right) \dd \vect{p} \dd z   \\
     &= -\frac1\beta  \int_{\R^n}  \sum_{j=1}^n  \sum_{i=1}^n \partial_{z_j} h_k\po z\pf _i\left( \rho_{\varphi}(z) M_\varphi(z)^{-1}\right)_{ij} \dd z \\
    &= \frac1\beta  \int_{\R^n}  \sum_{j=1}^n  \sum_{i=1}^n h_k\po z\pf_i \partial_{z_j}  \left( \rho_{\varphi}(z) M_\varphi(z)^{-1}\right)_{ij} \dd z\\
     &=  \frac1\beta  \int_{\R^n}  \sum_{i=1}^n  h_k(z)_i \sum_{j=1}^n \left(\partial_{z_j}  \left(M_\varphi(z)^{-1}\right)_{ji} +  \left(M_\varphi(z)^{-1}\right)_{ji} \partial_{z_j} \ln \rho_{\varphi}(z)\right) \rho_{\varphi}(z) \dd z\\
   & =  \frac1\beta  \int_{\R^n}  h_k(z)^\T  \cdot \left(\nabla\cdot  M_\varphi(z)^{-1} +  M_\varphi(z)^{-1}\cdot \nabla \ln \rho_{\varphi}(z)\right)  \rho_{\varphi}(z) \dd z
\end{align*}
which concludes the proof of \eqref{eq:Psi}.

\label{app:fdt}
\section{Details of the position-dependent 2-FDT}

The memory kernel is obtained through the Gram matrix $G_c$ of the function set and the scalar product $\sprod{C_k}{\mathcal{L} \xi_t }$. 
The Gram matrix of the set $\{c_k,{k\in\mathcal{J}}\}$ is related to the effective mass matrix,
\begin{align*}
    \label{eq:gram_matrix_ck}
     {G_{c_{k,k'}}} = \sprod{C_{k}}{C_{k'}} &=  \int_{\R^{D}} \int_{\R^{D}} \mathcal{O}_v(\vect{q},\vect{p})^\T \nabla h_k(\mathcal{O}_x(\vect{q})) ^\T  \nabla h_{k'}(\mathcal{O}_x(\vect{q}))  \mathcal{O}_v(\vect{q},\vect{p})   \rho_{eq}(\vect{q},\vect{p}) \dd \vect{q}\dd\vect{p} \\
     &=  \frac{1}{\beta} \sum_{j=1}^n  \sum_{i=1}^n\sum_{l=1}^n \int_{\R^n}  \partial_{z_j} h_{k}(z)_l  \partial_{z_i} h_{k'}(z)_l  \left(M_\varphi(z)^{-1}\right)_{ij}  \rho_{\varphi}(z) \dd z\\
     &= \frac{1}{\beta}\sprod{\nabla h_{k}}{M_\varphi^{-1}\cdot \nabla h_{k'}}_\varphi.
\end{align*}
Using the anti-hermitianity of $\mathcal{L}$, we have for $k \in \mathcal{J}$

\begin{eqnarray*}
\sprod{C_k}{\mathcal{L} \xi_t } &=&- \sprod{\mathcal{L} C_k}{\xi^v_t } \\
%%%%%%%% 
&=& -\sum_{i=1}^n \sum_{j=1}^n   \int_{\R^{D}} \int_{\R^{D}}  \partial_{z_j} h_k(\mathcal{O}_x(\vect{q}))_i  \mathcal{L}\mathcal{O}_{v,j} \xi^v_{t,i}(\vect{q},\vect{p})  \rho_{eq}(\vect{q},\vect{p}) \dd \vect{q}\dd\vect{p}  \\
&&- \sum_{i=1}^n \sum_{j=1}^n \sum_{l=1}^n  \int_{\R^{D}} \int_{\R^{D}}    \partial_{z_j}\partial_{z_l} h_k(\mathcal{O}_x(\vect{q}))_i\mathcal{O}_{v,l}(\vect{q},\vect{p})\mathcal{O}_{v,j}(\vect{q},\vect{p}) \xi^v_{t,i}(\vect{q},\vect{p})  \rho_{eq}(\vect{q},\vect{p}) \dd \vect{q}\dd\vect{p} .\\
\end{eqnarray*}
We can then introduce the identity in the form of $ 1- \mathcal{P}_{\mathcal{E}} +\mathcal{P}_{\mathcal{E}} $

\begin{eqnarray*}
\sprod{C_k}{\mathcal{L} \xi_t } &=&- \sprod{\mathcal{L} C_k}{\xi^v_t } \\
%%%%%%%%  
&=&-\sum_{i=1}^n \sum_{j=1}^n   \int_{\R^{D}} \int_{\R^{D}}  \partial_{z_j} h_k(\mathcal{O}_x(\vect{q}))_i  \left[ (1- \mathcal{P}_{\mathcal{E}} +\mathcal{P}_{\mathcal{E}} )\mathcal{L}\mathcal{O}_{v,j}\right] \xi^v_{t,i}(\vect{q},\vect{p})  \rho_{eq}(\vect{q},\vect{p}) \dd \vect{q}\dd\vect{p}  \\
&&- \sum_{i=1}^n \sum_{j=1}^n \sum_{l=1}^n  \int_{\R^{D}} \int_{\R^{D}}    \partial_{z_j}\partial_{z_l} h_k(\mathcal{O}_x(\vect{q}))_i\mathcal{O}_{v,l}(\vect{q},\vect{p})\mathcal{O}_{v,j}(\vect{q},\vect{p}) \xi^v_{t,i}(\vect{q},\vect{p})  \rho_{eq}(\vect{q},\vect{p}) \dd \vect{q}\dd\vect{p} \\
%%%%%%%% 
&=&-\sum_{i=1}^n \sum_{j=1}^n   \int_{\R^{D}} \int_{\R^{D}}  \partial_{z_j} h_k(\mathcal{O}_x(\vect{q}))_i \, \xi^v_{0,j}  \xi^v_{t,i}(\vect{q},\vect{p})  \rho_{eq}(\vect{q},\vect{p}) \dd \vect{q}\dd\vect{p}  \\
&&-\sum_{i=1}^n \sum_{j=1}^n   \int_{\R^{D}} \int_{\R^{D}}  \partial_{z_j} h_k(\mathcal{O}_x(\vect{q}))_i  \, f_b(\mathcal{O}_x(\vect{q}))_{j} \, \xi^v_{t,i}(\vect{q},\vect{p})  \rho_{eq}(\vect{q},\vect{p}) \dd \vect{q}\dd\vect{p}  \\
&&  - \sum_{i=1}^n \sum_{j=1}^n \sum_{l=1}^n  \int_{\R^{D}} \int_{\R^{D}}    \partial_{z_j}\partial_{z_l} h_k(\mathcal{O}_x(\vect{q}))_i\mathcal{O}_{v,l}(\vect{q},\vect{p})\mathcal{O}_{v,j}(\vect{q},\vect{p}) \xi^v_{t,i}(\vect{q},\vect{p})  \rho_{eq}(\vect{q},\vect{p}) \dd \vect{q}\dd\vect{p}, \\
\end{eqnarray*}
where we have used that
\begin{equation*}
    ( 1- \mathcal{P}_{\mathcal{E}} )
  \mathcal{L}\mathcal{O}_v= \xi^v_0 \quad \text{and}\quad \mathcal{P}_{\mathcal{E}}  \mathcal{L}\mathcal{O}_v=  f_b\circ \mathcal{O}_x.
\end{equation*}

By definition of the conditional variance of the noise~\eqref{eq:cond_variance_noise} and of the conditional expectation of the noise~\eqref{eq:cond_expec_noise} and \eqref{eq:cond_expec_noise_2}, we get
\begin{align*}
    \sprod{C_k}{\mathcal{L} \xi_t } =& -\int_{\R^n}  \sigma_\varphi(t,z)\cdot \nabla h_k(z)  \rho_{\varphi}(z) \dd z  \\
    &-\int_{\R^{n}} f_b(z)\cdot \nabla h_k(z) \cdot \bar{\xi}_t(z) \rho_{\varphi}(z) \dd z  \\ &-\int_{\R^n} \nabla^2 h_k (z) \,  \bar{\xi}^{\dot z}_t(z) \rho_{\varphi}(z) \dd z .
\end{align*}
Then a partial integration on the last term of the r.h.s. conclude the derivation of~\eqref{eq:kernel_FDT}.

\section{Projected correlation functions}
\label{app:proj_corr}
Correlation function of the noise with an observable $F(\vect{X})$ differs from the usual time correlation in that the noise is evolved with the orthogonal dynamics~\eqref{eq:ortho_dynamics}. Such projected correlations functions~\cite{Carof2014} are defined as
\begin{equation*}
    \sprod{F}{\xi_t} = \sprod{F}{e^{t(1-\mathcal{P}_{\mathcal{E}})\mathcal{L}} \xi_0}.
\end{equation*}
The noise follows from Eq.~\eqref{eq:GLEfgs} 
\begin{equation*}
    \xi_t\po \vect{X}\pf=  \Dp{\mathcal{O}(t)}{t}- f\po \mathcal O(t)\pf - \int_0^t \sum_{k} g_{k}(s) E_{k}(t-s)\dd s 
\end{equation*}
and the the projected correlation functions can be obtained from
\begin{equation*}
     \sprod{F}{\xi_t} =   \sprod{F}{\Dp{\mathcal{O}(t)}{t}- f\po \mathcal O(t)\pf}  - \int_0^t \sum_{k} g_{k}(s) \sprod{F}{E_{k}(t-s)}\dd s.
\end{equation*}
Notice that in particular a set of Volterra integral equations of the second kind for the memory kernel can be obtained. Indeed the coefficient of the memory kernel are projected correlation function, since for  $\ell\in\mathcal I$
\begin{equation*}
   \sum_{k}  G_{\ell,k} g_{k}(t) = \sprod{E_\ell}{\mathcal{L}\xi_t} = -\sprod{\mathcal{L}E_\ell}{\xi_t}
\end{equation*}
using the skew-symmetry of $\mathcal{L}$. This lead to~\cite{Givon2005}
\begin{equation*}
    \label{eq:volterra2nd}
  \sum_{k}  G_{\ell,k}g_{k}(t)    =-\left\langle \mathcal{L}E_\ell(0),\Dp{\mathcal{O}(t)}{t}\right\rangle+ \sum_{k}  f_{k} \sprod{\mathcal{L}E_\ell(0)}{E_{k}(t)} +\int_0^t \sum_{k} g_{k}(s) \sprod{\mathcal{L}E_\ell(0)}{E_{k}(t-s)} \dd s .
\end{equation*}

From a numerical point of view, we usually compute correlation function using a sum along trajectories sampled every $\Delta t$, \ie
\begin{equation*}
    \sprod{F}{\xi_t} = \sum_m F(\vect{X}(m\Delta t))^\T \xi_t(\vect{X}(m\Delta t)).
\end{equation*}
However, as the noise is evolved with the orthogonal dynamics, we need a relation between $\xi_t(\vect{X}(m\Delta t))$ and $\xi_{t+m\Delta t}(\vect{X})$. Writing $t=m\Delta t$, we write the noise at time $t+t_0$
\begin{equation*}
    \xi_{t+t_0}\po \vect{X}\pf=  e^{t\mathcal{L}}\Dp{\mathcal{O}(t_0)}{t}- e^{t\mathcal{L}}f\po \mathcal O(t_0)\pf - \int_0^{t+t_0} g\po s, \mathcal O(t+t_0-s)\pf\dd s.
\end{equation*}
Splitting the integral at $t$ and changing the variable $s$ of integration to $u=s-t$ leads to the required relation
\begin{equation*}
    \xi_{t+t_0}\po \vect{X}\pf= \xi_{t}\po \vect{X}(t_0) \pf - \int_0^{t_0} g\po u+t, \mathcal O(\vect{X}(t_0-u))\pf\dd u.
\end{equation*}
We emphasize that discretizing the integral gives a generalization of the algorithm for the computation of projected correlation function that was discussed in Refs.~\citenum{Carof2014,Lesnicki2016,Ayaz2022}.

\end{widetext}

\end{document}